\documentclass[a4paper, 12pt]{article}
\usepackage[top=2.8cm, bottom=2.8cm, outer=2.6cm, inner=2.6cm]{geometry}
\usepackage[onehalfspacing]{setspace}
\usepackage[format=hang, width=0.9\textwidth]{caption}
\usepackage{xspace}
\usepackage[shortcuts]{extdash}
\usepackage{authblk}
\usepackage{multirow}
\usepackage[hidelinks, english]{hyperref}
\usepackage[all]{hypcap}
\usepackage{cite}
\usepackage[separate-uncertainty=true]{siunitx}
\usepackage{orcidlink}

\newcommand\plotWidth{0.496\textwidth}
\newcommand\plotWidthWide{0.75\textwidth}

\newcommand{\pbpb}{Pb\==Pb\xspace}

\newcommand{\oo}{OO\xspace}
\newcommand{\po}{pO\xspace}
\newcommand{\nene}{Ne\==Ne\xspace}
\newcommand{\TeV}{\ensuremath{\mathrm{TeV}}\xspace}
\newcommand{\sigmainel}{\ensuremath{\sigma^{\mathrm{inel}}_{\mathrm{pp}}}\xspace}
\newcommand{\pt}{\ensuremath{p_{\mathrm T}}\xspace}
\newcommand{\xt}{\ensuremath{x_{\mathrm T}}\xspace}
\newcommand{\mpt}{\ensuremath{\langle p_\mathrm T\rangle}\xspace}
\newcommand{\mdndeta}{\ensuremath{\langle \mathrm{d} N_\mathrm {ch} / \mathrm{d}\eta \rangle}\xspace}
\newcommand{\sqrsn}{\ensuremath{\sqrt{s_\mathrm{NN}\,}}\xspace}
\newcommand{\sqrs}{\ensuremath{\sqrt{s\,}}\xspace}
\newcommand{\sqrsx}[1]{\ensuremath{\sqrt{s_{\mathrm{#1}}}}\xspace}
\newcommand{\gevc}{\ensuremath{{\mathrm{GeV}}/c}\xspace}
\newcommand{\raa}{\ensuremath{R_\mathrm{AA}}\xspace}
\newcommand{\rpo}{\ensuremath{R_\mathrm{pO}}\xspace}
\newcommand{\dndpt}{\ensuremath{\mathrm{d}N/\mathrm{d}\pt}\xspace}

\title{\Huge DNN predictions for pp reference \\ \pt spectra at unmeasured \sqrs}

\author{Maria A. Calmon Behling\thanks{\href{mailto:mcalmonb@physik.uni-frankfurt.de}{mcalmonb@physik.uni-frankfurt.de}} \,\orcidlink{0009-0009-0487-2555}, Mario Krüger\,\orcidlink{0000-0001-7174-6617}, \\ Jerome Jung\,\orcidlink{0000-0001-6811-5240} and Henner Büsching\,\orcidlink{0009-0009-4284-8943}}
\affil{\small Institut für Kernphysik, Goethe-Universität Frankfurt}
\date{}

\begin{document}

\maketitle
\thispagestyle{empty}
\begin{abstract}
Studies of the properties of the Quark–Gluon Plasma in high-energy heavy-ion collisions commonly facilitate proton–proton (pp) collisions at the same center-of-mass energy per nucleon pair as a reference measurement.
In this paper, a deep neural network-based approach for interpolating and extrapolating pp reference transverse-momentum spectra to unmeasured energies is presented.
The model is trained with ALICE data from LHC Runs 1 and 2 and provides predictions for center-of-mass energies relevant to LHC Run 3 and beyond.
\end{abstract}

\newpage
\section{Introduction}
The measurement of high transverse momentum (\pt) particles has proven to be very successful in characterizing properties of the hot and dense deconfined QCD matter, the Quark-Gluon Plasma (QGP), formed in high-energy heavy-ion collisions (AA). Commonly, this is studied by comparing the particle production in heavy-ion collisions with the particle production in proton–proton (pp) collisions at the same center of mass energy per nucleon pair (\sqrsn) by means of the nuclear modification factor (\raa)~\cite{PHENIX:2001hpc, PHENIX:2003djd, CMS:2012aa, ALICE:2012mj}.
In the absence of a suitable pp measurement, a pp reference is typically constructed from available measurements at other collision energies: 
Either a baseline \pt spectrum is scaled by a theoretically predicted energy dependence~\cite{CMS:2018yyx, ALICE:2021est, ATLAS:2022kqu}, or the pp reference is obtained directly by interpolating between two or more experimental measurements~\cite{PHENIX:2001hpc,CMS:2012aa,ATLAS:2016xpn,ALICE:2018hza,LHCb:2022tjh}.
The theory-driven approach is most reliable at high \pt, where perturbative QCD (pQCD) calculations are applicable, while at intermediate \pt, the energy dependence can only be derived from Monte Carlo event generators such as PYTHIA~\cite{Sjostrand:2014zea} with limited accuracy~\cite{ALICE:2015qqj}.
Data-driven interpolations, on the other hand, require the assumption of a functional form of the \sqrs dependence. 
At high \pt, functional forms assuming \xt-scaling of the spectra can be employed \cite{Brewer:2021tyv}, while at lower \pt, simple power-law parametrizations provide a reasonable description~\cite{ALICE:2018hza}. In \cite{shokr2021modeling}, however, it was shown that deep neural networks (DNNs) can be used to parametrize \pt spectra as a function of \sqrs. In contrast to the classical approaches, DNNs do not rely on assumptions about the energy dependence of particle production.

In this work, a data-driven, DNN-based method for constructing pp reference \pt spectra is presented, and its applicability to ALICE data at LHC energies is demonstrated.
\section{Dataset and data preparation}
The DNN is trained with inclusive charged-particle \pt spectra measured in pp collisions at five different LHC energies~\cite{ALICE:2022xip} (${\sqrs = 2.76,} $ $5.02,\,7,\,8,$ ${\mathrm{and} \, 13\,\mathrm{TeV}}$).
For each measured \sqrs, the dataset contains 46 \dndpt values with ${0.15<\pt/(\mathrm{GeV}/c)<10}$.

To optimize the performance of the DNN model in terms of computation time and prediction accuracy, a logarithmic transformation to the input (\pt, \sqrs), and output (\dndpt) values of the DNN is applied to similarize their scale. This is especially relevant for \dndpt, spanning several orders of magnitude.
An additional scaling of \dndpt by $1/\pt$ prior to the logarithmic transformation eases the extra\-polation of the training data towards low \pt values.
To account for uncertainties of the data, each point is varied multiple times (500) within the bounds of the corresponding systematic uncertainty of the measurement.
Subsequently, the data are randomly reordered and split into training (80\%) and validation (20\%) datasets to prevent overfitting. 
Additionally, the impact of low \pt data points during the training is increased by using sample weights.

\section{Model implementation}
\label{sec:ModelSetup}
A fully connected DNN is implemented in Tensorflow~\cite{tensorflow2015-whitepaper} and trained by minimizing the mean absolute error (MAE) between the data and DNN predictions using the Nadam optimizer~\cite{dozat2016incorporating}.
A simulation-based hyperparameter scan is performed to determine the most suitable architecture for accurate DNN predictions at unmeasured \pt and \sqrs values, employing PYTHIA-simulated data (v.8.3.06, Monash 2013 tune)~\cite{Skands:2014pea}.
For this study, only ALICE-equivalent data points in the simulation are used to train the model. 
Simulated data points at unmeasured \pt values and collision energies (selected in the range ${1.5 \leq \sqrs \leq  \SI{27}{TeV}}$) are then used to test the corresponding prediction accuracy. The upper limit of this energy range is motivated by the proposed upgrade of the LHC, the High-Energy Large
Hadron Collider (HE-LHC), which foresees a potential increase in the maximum center-of-mass energy to \sqrs = \SI{27}{TeV}~\cite{FCC:2018bvk}.
Six hyperparameters are varied in the scan: the number of hidden layers, the number of nodes per layer, the activation function, the initializer, the learning rate, and the batch size.
The optimal architectures are identified using Bayesian optimization~\cite{garnett_bayesoptbook_2023} as an efficient hyperparameter search strategy, implemented using the Keras Tuner~\cite{omalley2019kerastuner} package.
The overall best-performing architecture from the hyperparameter scan is chosen as the nominal architecture for the DNN model.

\section{Uncertainty estimation}
To estimate the uncertainties of the model predictions, two different contributions are considered: the aleatoric uncertainty from inherent randomness in the training process and the epistemic uncertainty from limited or sparse training data~\cite{MLUncertainties}.
Ensemble methods are used to determine these uncertainties.
The aleatoric uncertainty is quantified by training an ensemble of 20 DNN models with the nominal architecture, but different initialization values. 
The nominal prediction of the DNN is defined as the average over all predictions within this ensemble, and the aleatoric uncertainty as the corresponding standard deviation.
To quantify the epistemic uncertainty, a second ensemble of five DNNs is trained with different architectures (the five best-performing architectures from the hyperparameter scan). The epistemic uncertainty for a given prediction is then calculated as the RMS of the deviations between the DNN model predictions in the ensemble to that of the nominal architecture.
Finally, the total model uncertainty is given by the quadratic sum over the aleatoric and epistemic uncertainties.

\section{Performance evaluation}
\label{sec:ClosureTest}

\begin{figure}[t]
	\center
	\includegraphics[width=\plotWidth]{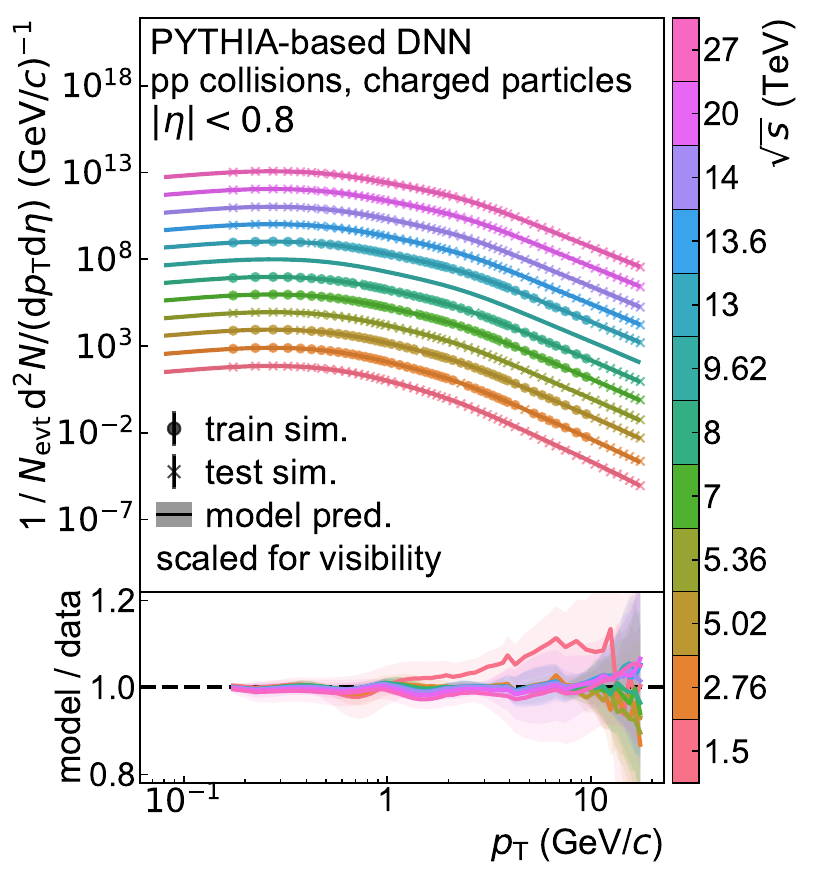}
	\includegraphics[width=\plotWidth]{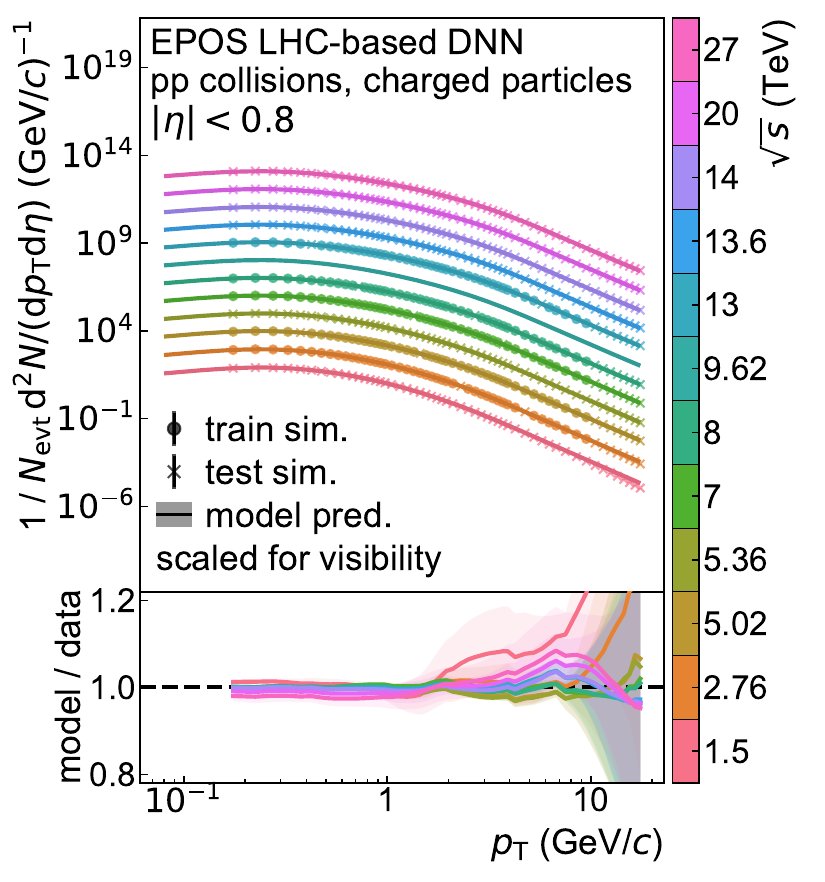}
	\caption{\pt spectra simulated with PYTHIA (left) and EPOS LHC (right) together with the corresponding predictions of the PYTHIA- and EPOS LHC-based DNNs. 
    }
	\label{fig:MC_based_fit}
\end{figure}

The extrapolation performance of the DNN architecture determined with PYTHIA ('PYTHIA-based DNN') in the hyperparameter scan is evaluated using an independent dataset of EPOS LHC-simulated data~\cite{Pierog:2013ria}, as the \pt spectra in PYTHIA and EPOS LHC are different.
Again, only ALICE-equivalent data points in the simulation are used to train the DNN ensemble ('EPOS LHC-based DNN').
\autoref{fig:MC_based_fit} shows the corresponding DNN predictions within a range of ${\sqrs=1.5}$ to \SI{27}{TeV}. Both DNNs achieve an excellent description of the training data.
The interpolation performance within the energy range of the training data is equally precise.
However, the prediction accuracy decreases with energy difference to the training data: The largest deviations are observed at ${\sqrs=\SI{1.5}{TeV}}$, illustrating that the DNN extrapolation is more challenging towards lower unmeasured energies than to higher energies.
Furthermore, the high-\pt extrapolation accuracy (${\pt > \SI{10}{\gevc}}$) worsens with increasing \pt. 
Since the PYTHIA-simulated data was used to select the model architectures, the PYTHIA-based DNN is slightly more accurate than the EPOS LHC-based DNN. However, both provide very good predictions across the whole studied \sqrs range, and deviations from the test data are generally covered by the assigned model uncertainties. 

\section{Model application to ALICE data}
\begin{figure}[t]
	\center
	\includegraphics[width=\plotWidth]{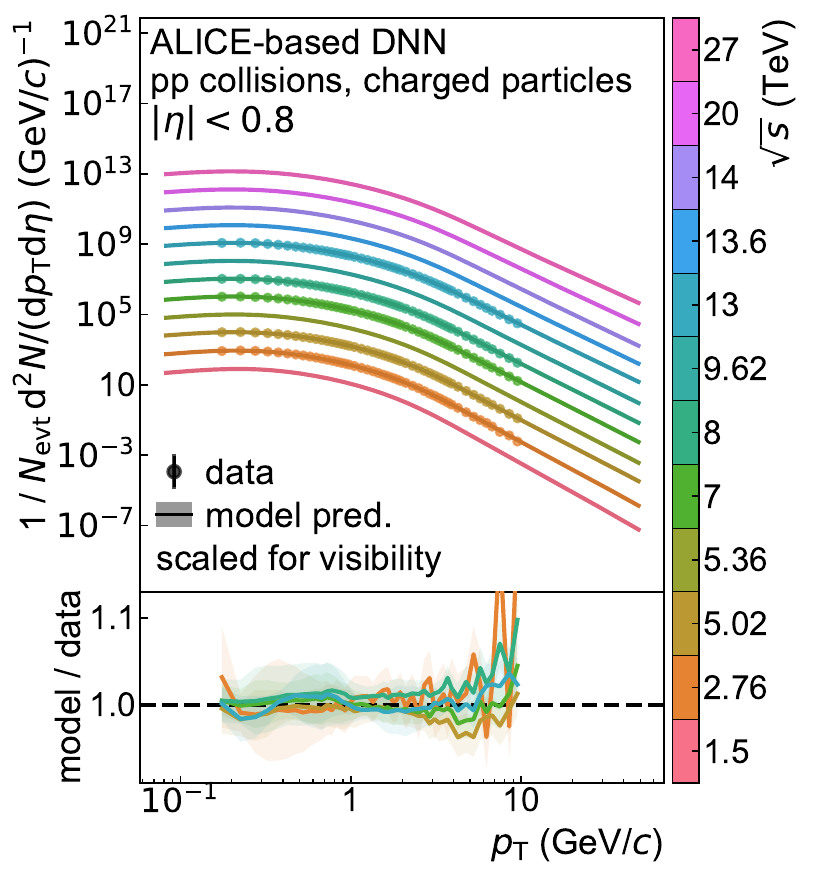}
	\includegraphics[width=\plotWidth]{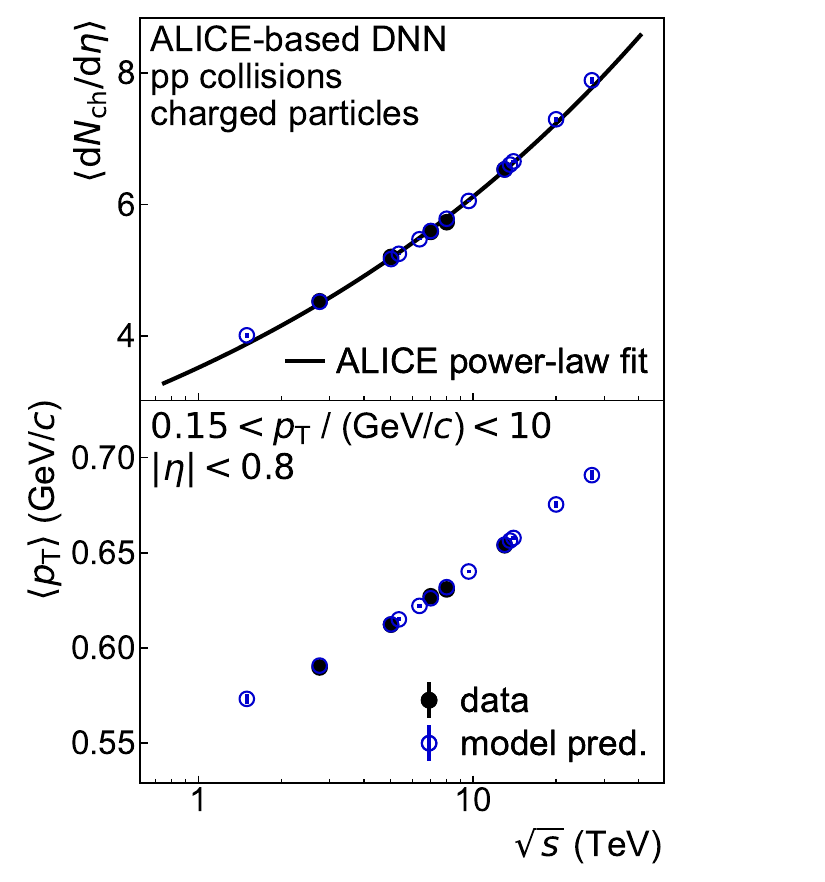}
	\caption{\pt spectra measured by ALICE~\cite{ALICE:2022xip} together with the corresponding predictions of the ALICE-based DNN model (left) and their \mdndeta and \mpt as a function of \sqrs (right). 
    }
	\label{fig:alice_based_fit}
\end{figure}

After selecting the model architecture based on the PYTHIA dataset and validating the performance with the EPOS LHC dataset, the final DNN ensemble ('ALICE-based DNN') is trained on the ALICE dataset~\cite{ALICE:2022xip}.
Additional constraints on the DNN are imposed by extending the training dataset with supplementary high-\pt data points sampled from a power-law parametrization ($\propto 1/\pt^{n}$~\cite{Cronin:1973fd,British-ScandinavianISR:1973vpp,Hagedorn:1983wk,Wong:2015mba}) up to ${\pt = 50\,\gevc}$ to improve the DNN extrapolation accuracy in the high \pt limit of the ALICE measurement. 
\autoref{fig:alice_based_fit} (left) shows the predicted \pt spectra of the ALICE-based DNN at training energies and unmeasured energies within ${\sqrs = 1.5-\SI{27}{TeV}}$ and ${0.1 < \pt / (\gevc) < 50}$.
An excellent parametrization of the ALICE data is achieved across the whole \sqrs and \pt range. Deviations from the data are well covered by the corresponding DNN uncertainties.
\autoref{fig:alice_based_fit} (right) shows the derived \mdndeta and \mpt of both the DNN predicted and ALICE-measured \pt spectra as a function of \sqrs, limited to the measured \pt range.
In both cases, the predicted values align well with the ALICE training data. Furthermore, the predicted values of \mdndeta at unmeasured energies are consistent with an expected power-law behavior (${\mdndeta \propto \sqrs^{b}}$)~\cite{ALICE:2022kol}.
\section{Construction of pp reference \pt spectra}
\begin{figure}[t]
	\center
    \includegraphics[width=\plotWidthWide]{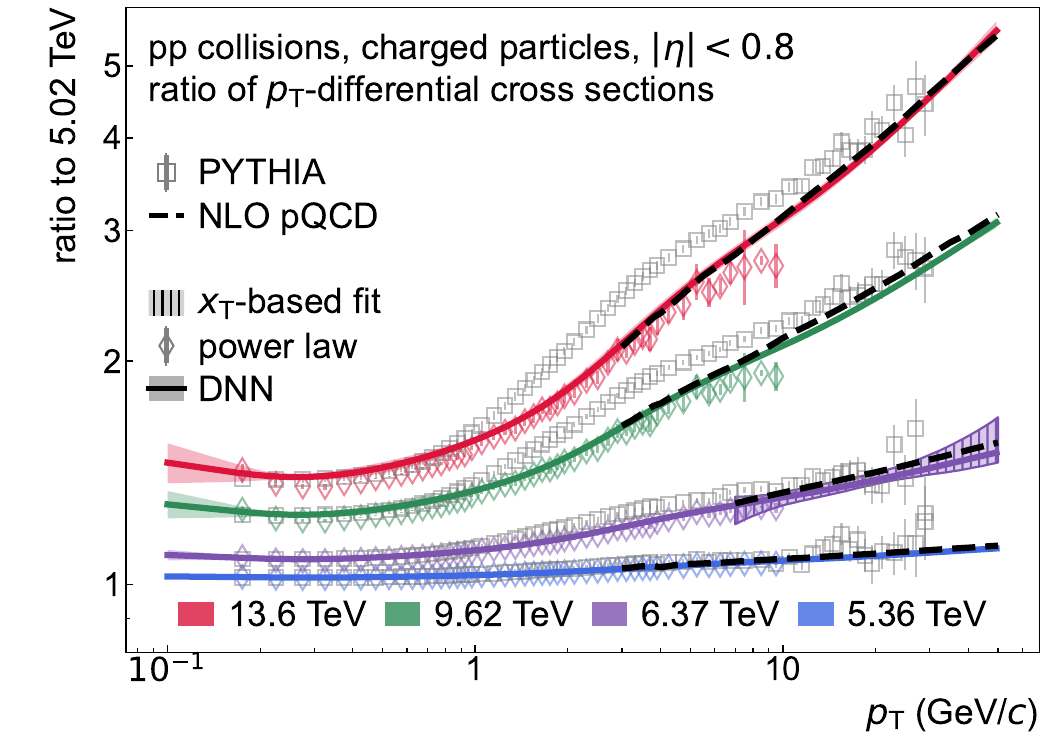}
    \caption{Ratios of \pt-differential cross sections at different energies to ${\sqrs= \SI{5.02}{TeV}}$ as predicted by the DNN, together with NLO pQCD calculations, PYTHIA simulations, and functional interpolations, summarized in \autoref{tab:XSecPred}.}
	\label{fig:moneyPlot}
\end{figure}
The ALICE-based DNN predictions can now be used to construct 'pp reference \pt spectra' for the calculation of \raa. 
As the event normalization of the training data and the pp reference spectra are different, this needs to be corrected for.
The approach used in the following is based on the ratios of DNN predictions and a correctly normalized spectrum at a baseline energy:
The pp reference \pt spectrum at a target energy \sqrsx{T} is constructed from the baseline spectrum at an energy \sqrsx{B}, scaled by the corresponding ratio of ALICE-based DNN predictions of energies \sqrsx{T} and \sqrsx{B}, as the ratio of the two DNN predictions at different \sqrs is independent of the overall event normalization.
In order to scale \pt-differential cross sections, an additional normalization by the corresponding ratio of total inelastic pp cross sections $\sigmainel(\sqrsx{T})/\sigmainel(\sqrsx{B})$ must be applied. 
The values and uncertainties of \sigmainel are determined using a parametrization of world data~\cite{Loizides:2017ack}.
\\
In \autoref{fig:moneyPlot}, the DNN-predicted ratios of \pt-differential cross sections to the baseline energy ${\sqrs = \SI{5.02}{TeV}}$ are shown, relating energies relevant for LHC~Run~3 measurements with the pp reference energy of the LHC~Run~2 \pbpb data:
${\sqrs = \SI{5.36}{TeV}}$ as pp reference energy for \oo, \nene, and \pbpb collisions, ${\sqrs = \SI{6.37}{TeV}}$ as pp reference energy for the earlier planning of \oo collisions~\cite{Citron:2018lsq}, ${\sqrs = \SI{9.62}{TeV}}$ as pp reference energy for \po collisions and ${\sqrs = \SI{13.6}{TeV}}$ as the nominal pp collision energy at LHC.

To compare the DNN approach to traditional theory-based (PYTHIA, NLO pQCD) and data-driven (power-law and \xt-based) interpolation methods for the construction of a pp reference \pt spectrum, \autoref{fig:moneyPlot} also includes the ratios of the corresponding predictions for these methods. Details are listed in \autoref{tab:XSecPred}.
For the power-law interpolation, the ALICE data are parametrized with $a \cdot \sqrs^{b}$ separately for each \pt interval, as applied in~\cite{ALICE:2018hza}.
While at ${\sqrs = \SI{5.36}{\TeV}}$ all methods provide similar predictions, their differences become more pronounced as \sqrs increases. At low and intermediate \pt, only the PYTHIA, power-law, and DNN approaches can provide predictions. These three methods are largely consistent within uncertainties up to ${\pt \approx \SI{0.9}{\gevc}}$. 
Beyond this region, PYTHIA begins to deviate from the power-law and DNN predictions, with the deviation becoming more pronounced at higher energies. This behavior has been previously observed in~\cite{ALICE:2015qqj}.
While the power-law approach shows fluctuations which arise from the individual interpolation of \pt intervals, it remains highly consistent with the DNN approach up to ${\pt \approx \SI{5}{\gevc}}$ and becomes less reliable at high \pt, as expected~\cite{ALICE:2021wim}. 
In this region, \xt-based interpolations are better suited to describe the energy dependence and are compatible with the DNN predictions. 
\begin{table}[t]
\centering
\begin{tabular}{lccc}
\hline\hline
\multirow{2}{*}{Method} & \sqrs & \pt range & \multirow{2}{*}{Reference} \\
     & (TeV)  &  (\gevc)  & \\
\hline\hline
PYTHIA simulations & 5.36, 9.62, 6.37, 13.6 & [0.15 , 30] & \cite{Skands:2014pea} \\
\multirow{2}{*}{NLO pQCD calculations}
  & 6.37 & [7.00 , 50] &  \cite{Brewer:2021tyv} based on \cite{Aversa:1988vb} \\
  & 5.36, 9.62, 13.6 & [3.00 , 50] &   \cite{Jonas:2026yoz} based on \cite{incnlowebsite,INCNLO1,INCNLO2}\\
\hline
\xt-based interpolation  & 6.37 & [7.00 , 50] & \cite{Brewer:2021tyv} \\
power law interpolation  &  5.36, 9.62, 6.37, 13.6 & [0.15 , 10] & This work \\
DNN interpolation  &  5.36, 9.62, 6.37, 13.6 & [0.10 , 50] & This work \\
\hline\hline
\end{tabular}
\caption{Methods for predicting pp reference cross sections.}
\label{tab:XSecPred}
\end{table}
For all energies, the DNN predictions align well with the NLO pQCD expectations over the entire overlap region of ${3< \pt / (\gevc) < 50}$.

\begin{figure}[t]
	\center
    \includegraphics[width=\plotWidthWide]{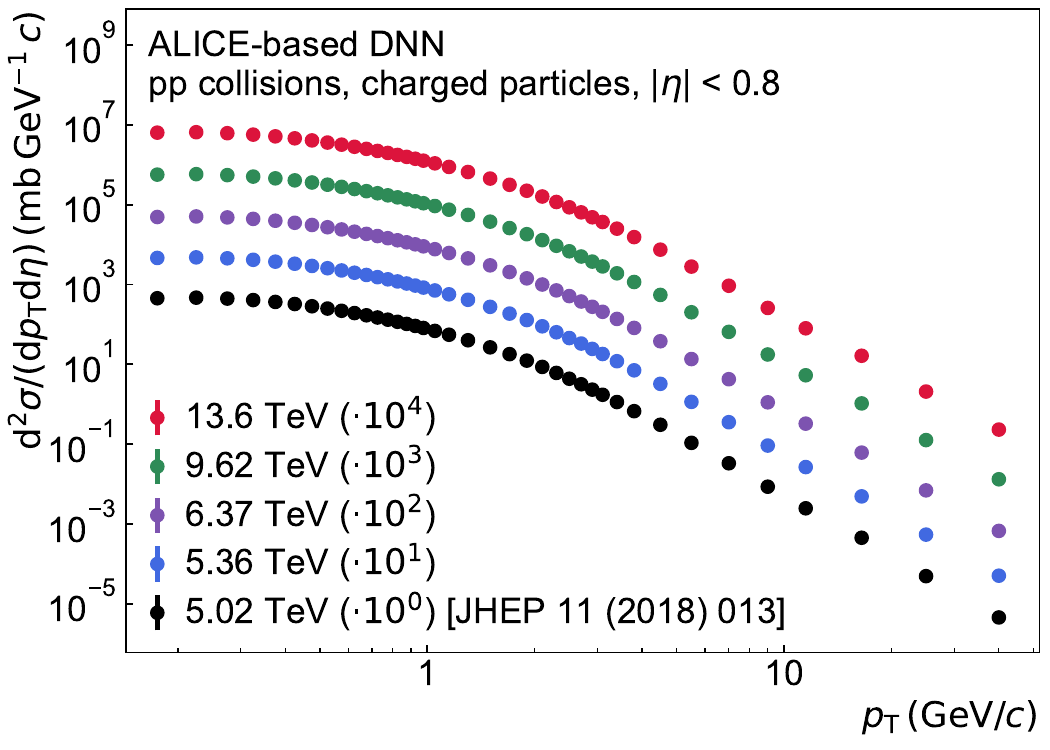}
    \caption{pp reference \pt-differential cross sections at different energies constructed from the DNN-predicted cross section ratios to the baseline energy ${\sqrs = \SI{5.02}{TeV}}$ and a corresponding ALICE measurement~\cite{ALICE:2018hza}.}
	\label{fig:ppRef}
\end{figure}

Finally, the pp reference \pt-differential cross sections of the four energies, constructed by multiplying the DNN-predicted ratios (shown in \autoref{fig:moneyPlot}) with a corresponding baseline measurement at ${\sqrs = \SI{5.02}{TeV}}$ from~\cite{ALICE:2018hza}, are shown in \autoref{fig:ppRef}.
The DNN-based pp reference at ${\sqrs = \SI{9.62}{TeV}}$ enables the calculation of \rpo, facilitating the recent LHC \po data~\cite{ATLAS:2026xcm}, for which no corresponding pp measurement exists.
While the study in this paper was performed for inclusive charged particles, the DNN approach can be extended to individual particle species, as well.
\section{Summary}
In this paper, a DNN-based method for constructing pp reference \pt spectra at unmeasured \sqrs is presented.
The DNN is trained with inclusive charged-particle \pt spectra measured by the ALICE collaboration in pp collisions at five different LHC energies.
A hyperparameter scan based on PYTHIA simulations is performed to determine the most suitable architecture for accurate DNN predictions within and beyond the LHC energy range. 
The extrapolation performance of the DNN architecture is evaluated using an independent dataset of EPOS LHC simulations.
The final DNN model is trained on the ALICE dataset and then used to construct pp reference \pt spectra for calculating \raa at selected energies.
The DNN approach compares well with PYTHIA at low \pt, with a power-law approach at low and intermediate \pt, and with NLO pQCD calculations as well as \xt-based interpolations at high \pt.
It provides continuous predictions across a wide \pt range without making assumptions about the physical processes underlying the spectra's energy dependence.
\section*{Acknowledgements}
The authors thank Hannah Bossi for the valuable input and discussions, and Nicolas Strangmann and Florian Jonas for providing the NLO pQCD calculations.
\newpage
\bibliographystyle{utphys}
\bibliography{bibliography}
\end{document}